\documentclass[11pt,a4paper]{article}

\usepackage[english]{babel}
\usepackage[left=0.8in,right=0.8in,top=1.0in,bottom=1.2in]{geometry}
\usepackage[labelfont=bf]{caption}
\usepackage{amsfonts}
\usepackage{amsmath}
\usepackage{amssymb}
\usepackage{amsthm}
\usepackage{pdfpages}
\usepackage{hyperref}
\usepackage{multirow}
\usepackage{graphicx}
\usepackage{mathtools}
\usepackage{braket}
\usepackage{cite}
\usepackage{array}
\usepackage{arydshln}
\usepackage{epstopdf}
\usepackage{soul}
\numberwithin{equation}{section}
\numberwithin{figure}{section}
\numberwithin{table}{section}

\newcommand{\be}{\begin{equation}}
\newcommand{\ee}{\end{equation}}
\newcommand{\bea}{\begin{eqnarray}}
\newcommand{\eea}{\end{eqnarray}}  

\newcommand{\gsim}{\lower.7ex\hbox{$\;\stackrel{\textstyle>}{\sim}\;$}}
\newcommand{\lsim}{\lower.7ex\hbox{$\;\stackrel{\textstyle<}{\sim}\;$}}
\newcommand{\cO}{{\mathcal O}}

\newcommand{\abs}[1]{\left| #1 \right|}

\makeatletter
\g@addto@macro\bfseries{\boldmath}
\makeatother

\begin{document}

\begin{flushright}
ZU-TH 37/22\\
MITP-22-068
\end{flushright}

\begin{center}
\vspace{0.7cm}
{\Large\bf Simplified models of vector $SU(4)$ leptoquarks at the TeV}\\[1.0cm] 
Riccardo Barbieri$^a$, Claudia Cornella$^b$, Gino Isidori$^c$\\[0.2cm]
{\em ${}^a$Scuola Normale Superiore, Piazza dei Cavalieri 7, 56126 Pisa, Italy}\\
{\em ${}^b$ PRISMA$^+$ Cluster of Excellence $\&$ MITP,  Johannes Gutenberg  Universit\"at, 55099 Mainz, Germany}\\
{\em ${}^c$Physik-Institut, Universit\"at Z\"urich,  8057 Z\"urich, Switzerland}
\vspace{0.5 cm}
\end{center}

\centerline{\large\bf Abstract}
\begin{quote}
Assuming confirmation of the anomalies in semi-leptonic $B$-decays, their explanation in terms of the exchange of a massive vector leptoquark field, $U_\mu^a$, of charge 2/3, appears to require the inclusion of $U_\mu^a$ in the vector multiplet of the adjoint of Pati-Salam $SU(4)$,  $\mathcal{G}_\mu^A$,  as well as the introduction of vector-like fermions, $F_j$, in the fundamental of  $SU(4)$.

We consider simplified models characterised by the nature of the  $SU(4)$ symmetry (global or local), by the number of vector-like fermions, and by
the couplings of the Higgs boson to SM fermions (direct or induced by mixing with the vector-like fermions). In all cases, we implement a minimal breaking of a $U(2)^n_f$ flavour symmetry, with a single motivated exception. We then perform a global fit including the main observables sensitive to exchanges of the $\mathcal{G}_\mu^A$ at tree level and in loops dominated by logs insensitive to the UV completion. 
\end{quote}

\section{Introduction and motivations}

The intriguing deviations from the Standard Model (SM) appearing in  
neutral-current~\cite{LHCb:2014vgu,LHCb:2017avl,LHCb:2019hip,LHCb:2021trn} and charged-current~\cite{BaBar:2012obs,BaBar:2013mob,Belle:2015qfa,LHCb:2015gmp,LHCb:2017smo,LHCb:2017rln} 
semileptonic $B$-meson decays, if confirmed by experiments in progress, would point towards the existence of new physics close to the TeV scale, and hopefully shed some much-needed light on the flavour problem of the SM. This amply justifies the interest in the subject in the literature, with many different attempts put forward to explain these  ``B-anomalies". Among them one emerging possibility is that the anomalies are due to the exchange of a vector leptoquark of charge $2/3$, $U_\mu^a$ \cite{Barbieri:2015yvd} (see also Refs.~\cite{Alonso:2015sja,Calibbi:2015kma,Buttazzo:2017ixm}), 
for which several UV-complete models have been already suggested~\cite{Barbieri:2016las,DiLuzio:2017vat,Barbieri:2017tuq,Bordone:2017bld,Greljo:2018tuh,Cornella:2019hct,Fuentes-Martin:2020bnh,Fuentes-Martin:2022xnb}.

Given this framework, generally speaking, two features seem unavoidable:
\begin{itemize}
\item[i.] The inclusion of $U_\mu^a$ in the vector multiplet, $\mathcal{G}_\mu^A$, of the adjoint of Pati-Salam $SU(4)$  with at least the first two generations of the Standard Model (SM) fermions not directly coupled to $\mathcal{G}_\mu^A$. Here we assume that at least some of the three generations of SM fermions, $f_i$, couple to the $\mathcal{G}_\mu^A$ by mixing with  a suitable number of vector-like fermions, $F_j$, in the fundamental of $SU(4)$.

\item[ii.] An approximate flavour symmetry to allow for a relatively low mass for the vectors. To make contact with the observed pattern of quark and lepton masses and of quark mixings, we  implement a minimal breaking of a $U(2)^n_f$ flavour symmetry, with a single motivated exception.
\end{itemize}

In this work, we propose a few simplified models that capture the essence of the vector leptoquark explanation of the anomalies and we analyse their consistency with the observables affected by $\mathcal{G}_\mu^A$ exchanges at tree level or in loops dominated by IR logs. While this is a seemingly step backward with respect to offering a full UV-complete model, we believe that this may help identify the proper direction for model building, as well as better appreciate the role of the expected experimental progress in different observables. To this end the tie with the observed flavour structure of the SM -- which we shall implement differently in the different models  -- plays, in our view, a decisive role.

As said, we chose not to constrain ourselves into a specific UV-complete model. Nevertheless, broadly speaking, we have in mind at least two possibilities for the $\mathcal{G}_\mu^A$. They can be $\rho$-like states associated with a $SU(4)$ global symmetry of some strong dynamics at the TeV scale, with standard color $SU(3)$ gauged inside $SU(4)$~\cite{Barbieri:2016las,Barbieri:2017tuq}. Alternatively,  $SU(4)$ can be part of a fully gauged $SU(4)\times SU(3)^\prime$, suitably included into a larger gauge group, with $SU(4)\times SU(3)^\prime$ broken to the diagonal standard colour $SU(3)$~\cite{DiLuzio:2017vat,Bordone:2017bld}.
A model formulated in more than four dimensions can allow a bridge between these two scenarios~\cite {Fuentes-Martin:2020bnh,Fuentes-Martin:2020pww}. It is also conceivable, though not of concern in this work, that these models be extended to include as well a composite picture of ElectroWeak symmetry breaking, as in Refs.~\cite{Barbieri:2017tuq,Fuentes-Martin:2020bnh,Fuentes-Martin:2022xnb}.

The paper is organised as follows. In Section 2 we define four different simplified models. In Section 3, for each model we determine the couplings of the massless fermions to the $\mathcal{G}_\mu^A$ before EW symmetry breaking, as well as their Yukawa couplings. Section 4 contains the phenomenological analysis. A discussion of the results is summarised in Section 5.

\section{Definition of the models}
A common element of the models we consider is a minimal set of  $J$ vector-like fermions,
\begin{equation}
F_j= \left(\begin{array}{c}
Q^a_j \\
L_j
\end{array}\right)\,, \qquad j=1 \dots J\,, 
\label{Fi}
\end{equation}
transforming in the fundamental of $SU(4)$. The apex $a$ is a color index, and $Q^a_j$ and $L_j$ are doublets under the standard $SU(2)_L$ gauge group, commuting with $SU(4)$. The $F_j$ have a universal mass term, $\mathcal{L} _M$, 
and a mass-mixing with the standard fermions $f_i$, $\mathcal{L}_{\rm mix}$: 
\begin{equation}
\mathcal{L}_m = \mathcal{L} _M+ \mathcal{L}_{\rm mix}\,,
\label{Lm}
\end{equation}
and they may enter in the Yukawa couplings to the Higgs scalar in $\mathcal{L}_Y$.
Both $\mathcal{L}_m$ and $\mathcal{L}_Y$ have to be invariant under the SM gauge group. The natural presence of further vector-like fermions in the fundamental of $SU(4)$ and transforming as singlets under $SU(2)_L$ does not play any significant phenomenological role, as we shall see in the following.

The four models that we consider (see Table~\ref{Tab:models}) are classified according to the following properties:
\begin{itemize}
\item The coupling of the Higgs boson to the standard fermions $f_i$ is direct or arises only after their mixing with the vector-like fermions $F_j$;

\item $SU(4)$ is a global or a local symmetry. In the latter case $SU(4)$ acts not only on the  $F_j$ but also on the third family of standard quarks and leptons, extended to include a right-handed neutrino and organised in the usual Pati-Salam 4-plets.

\end{itemize}

In all cases, with a single motivated exception, we implement in both $\mathcal{L} _Y$ and $\mathcal{L} _{\rm mix}$ a minimal breaking of a $U(2)^n_f$ flavour symmetry acting on the first two generations of the $f_i$~\cite{Barbieri:2011ci,Barbieri:2012uh,Blankenburg:2012nx}  and, depending on the model, extended to the $F_j$.
\begin{table}[t]
\begin{center}
\renewcommand{\arraystretch}{1.2} 
\begin{tabular}{c||c|c|c|c}
Model & {Direct SM Yukawa}  & $SU(4)$ gauged & min. $U(2)^n_f$ breaking & $J$  \\ \hline\hline
1& yes & no & yes&2\\ \hline
2&yes&yes&yes&1 \\ \hline
3&yes&yes&no&2 \\ \hline
4&no&no&yes& $3(\times 2)$ \\ \hline
\end{tabular}
\end{center}
\caption{Main features of the four simplified models considered.
\label{Tab:models}}
\end{table}
%
\subsection{Model 1}
In this model, the SM fermions do not couple directly to the $\mathcal{G}_\mu^A$, while the 
two vector-like fermions $F_j$ are assumed to interact universally, in a $SU(4)$ invariant way, with the $\mathcal{G}_\mu^A$:

\begin{equation}
\mathcal{L}_{\rm int}= g_U \mathcal{G}_\mu^A J^{\mu\, A} ~=~ g_U \left[\frac{1}{\sqrt{2}} \left( U_\mu^a J_U^{\mu\,a}+{\rm h.c.}\right)  +  G^{\hat{a}}_\mu J_{G}^{\mu\,\hat{a}} +\frac{1}{2 \sqrt{6}} \, X_\mu J_{X}^\mu \right] \,.
\end{equation}
Here we have introduced, together with the leptoquark $U_\mu^a$, the coloron $G^{\hat{a}}_\mu$ and the $B$--$L$ vector $X_\mu$.
The  currents, written in terms of the components of the $F_j$, are
\begin{equation}
{J}^{\mu\,a}_U  ~=~     \bar Q_j^a   \gamma_{\mu}  L_j\,,\quad
J_{G}^{\mu\,\hat{a}} ~= ~   \bar Q_j  T^{\hat{a}} \gamma_{\mu}  Q_j \,,\quad
{J}^\mu_{X} ~= ~ \bar Q_j\gamma_{\mu}  Q_j  {- 3}  \bar L_j \,   \gamma_{\mu} L_j \,.
\label{SU4currents}
\end{equation}
with $j=1,2$. The flavour symmetry acting on the first two generations of chiral fermions $f_i$  is
\begin{equation}
U(2)^5_f=U(2)_q\times U(2)_u\times U(2)_d\times U(2)_l\times U(2)_e,
\end{equation}
where we used the standard notation for the irreducible representations of the SM gauge group. 
Following~\cite{Barbieri:2011ci,Barbieri:2012uh,Blankenburg:2012nx}, we define as
``minimal'' the case where the breaking of  $U(2)^5_f$ occurs only via 
\begin{itemize}
\item  the (leading) spurion doublets,
\begin{equation}
V_q \sim 2_q\,, \qquad V_l  \sim 2_l\,,
\end{equation}
whose natural size is set by the $32$ mixing in the CKM matrix ($|V_{cb}| \approx 4 \times 10^{-2}$);
\item  the (subleading) spurion bi-doublets 
\begin{equation}
\Delta_u  \sim (2_q, \bar{2}_u)\,,  \quad \Delta_d  \sim (2_q, \bar{2}_d)\,, \quad \Delta_e  \sim (2_l, \bar{2}_e)\,,
\end{equation}
each with their two eigenvalues of order of magnitude similar to the two lightest quark and lepton masses relative to the third one. \end{itemize}

The SM fermions couple directly to the Higgs field.
The $U(2)^5_f$ breaking structure implies that the $3\times 3$ Yukawa matrices have the form
\begin{equation}
 \mathcal{L}_Y^u =  y_t H (\bar{q}, \bar{q}_3)
\left(\begin{array}{c:c}
 \Delta_u  & x_tV_q \\\hdashline
 0 & 1
\end{array}\right)u_R \equiv H \bar{q}_L\hat{Y}_uu_R\,,
\label{Yuka}
\end{equation}
with $x_t =\cO(1)$, 
and similarly for $ \mathcal{L}_Y^{d,e}$. After $SU(4)$ breaking,  $\mathcal{L}_m =\mathcal{L}_{m}^q+ \mathcal{L}_{m}^l$ can be decomposed as 
\begin{equation}
\mathcal{L}_{m}^q = M[\bar{Q}_{j_L}+\alpha_j \bar{q}_{3} +\beta_j (\bar{q} V_q)]Q_{j_R}\,,
\label{yandm}
\end{equation}
with $\alpha_j, \beta_j=\cO(1)$,  and similarly for $\mathcal{L}_{m}^l$. 
In both Eq.~\eqref{Yuka} and Eq.~\eqref{yandm} $V_q$ is contracted with the doublet component of $q$ under $U(2)_q$.
Note that the minimality of the $U(2)^5_f$ breaking forbids any mixing of the two light generations with possible vector-like $SU(2)_L$-singlet fermions, thus justifying not having included the latter in the first place.
\subsection{Model 2}
At variance with Model 1, in Model 2 $SU(4)$ is gauged and acts also on the third family of SM quarks and leptons, embedded in Pati-Salam 4-plets with the addition of a right-handed neutrino. We also assume a single family of vector-like fermions charged under $SU(2)_L$ and $SU(4)$. The relevant part of the $SU(4)$ currents in Eq.~(\ref{SU4currents}) receives  extra pieces:
\begin{align}
{J}^{\mu\,a}_U  ~=~  &   \bar Q^a_j  \gamma_{\mu}  L_j  +\bar q^a_{3}  \gamma_{\mu}  l_{3}\,,  \\
\label{currentU}
J_{G}^{\mu\,\hat{a}} ~= ~  &    \,  \bar Q_j   T^{\hat{a}} \gamma_{\mu}  Q_j +\bar q_{3}  T^{\hat{a}} \gamma_{\mu}  q_{3}\,,    \\
{J}^\mu_{X} ~= ~&  \bar Q_j \gamma_{\mu}  Q_j  {- 3}  \bar L_j\,   \gamma_{\mu} L_j+ \bar q_{3}  \gamma_{\mu}  q_{3} {- 3}  \bar l_{3} \,   \gamma_{\mu} l_{3} \,,  
\label{currentX}
\end{align}
where we include explicitly only the left-handed $SU(2)_L$ doublets.\footnote{~In Models 2 and 3 it is natural 
to expect also couplings of the heavy vectors to right-handed third-generation chiral fermions~\cite{Bordone:2017bld}. The phenomenology 
of such right-handed currents has been discussed in detail in Refs.~\cite{Cornella:2019hct,Fuentes-Martin:2019mun}. For the sake of minimality,  
we do not consider these couplings here. To this purpose, we note that even if right-handed chiral fermions
are charged under $SU(4)$, their couplings to the heavy vectors can be suppressed via 
a mass mixing with vector-like fermions which are not charged under $SU(4)$. }
We left the index $j$ for later purpose, although in this case 
there is a  single family of vector-like fermions, dubbed $j=2$ for ease of notation. Unlike in Model~1, in this case, the universality of the couplings in these currents is dictated by gauge invariance.

The structure of both Yukawa coupling and vector-like mass terms is like in Model 1. Note, however, that in this case, we need $SU(4)$-breaking 
terms not only in $\mathcal{L}_m$ but also in the Yukawa coupling:  the mixing between 
light and third generations in Eq.~\eqref{Yuka} breaks explicitly the $SU(4)$ gauge symmetry. 
This term can be viewed as the effective result of a $SU(4)$-conserving Yukawa interaction between vector-like fermions 
and the Higgs, after integrating out the heavy fermions. 
\subsection{Model 3}
In Model 3 there is no change with respect to Model 2 either in the gauge structure or in the Yukawa couplings. 
However, we enlarge the matter field content with a second family of vector-like fermions ($j=1,2$).
This allows us to extend the flavour symmetry to $U(2)^5_f\times U(2)_F$,
and to introduce a non-minimal breaking of  $U(2)^5_f$ via the bi-doublets 
\begin{equation}
\Delta^q = (2_q, \bar{2}_F)\,, \qquad \Delta^l=(2_l, \bar{2}_F)\,.
\end{equation}
The latter control the mixing between light families and vector-like fermions via 
\begin{equation}
\mathcal{L}_{m}^q = M  \left[ \bar{Q}_{j_L} +\alpha_j \bar{q}_{3} +
\bar{q}_{i} \Delta^q_{ij} \right] Q_{j_R}  \,,
\label{Lqm}
\end{equation}
and similarly in $\mathcal{L}^q_m$. 
This will have important consequences for the alignment of the Yukawa couplings, which are not generic, with the mass eigenstates.
\subsection{Model 4}
As we are going to see, in none of the previous models there is a fixed orientation of the vector-like--SM-fermion mixing with respect to the
up or down  SM Yukawa couplings. This introduces an intrinsic uncertainty of the order of the
Cabibbo-Kobayashi-Maskawa matrix,  $V_{\rm CKM}$, in this relative orientation. 
To avoid this feature, we consider a model in which, as in Ref.~\cite{Barbieri:2017tuq}, there is a doubling of the three vector-like $SU(4)$ multiplets (with the same weakly gauged quantum numbers)
\begin{equation}
F^u_i=(Q^u,L^\nu)_i\,,\quad 
F^d_i=(Q^d,L^e)_i\,,\quad\quad i=1,2,3\,,
\end{equation}
and assume no direct coupling between the $SU(2)_L$-charged  SM fermions and the Higgs.
We further assume that:
\begin{itemize}
\item
 As in Model 1, only the vector-like fermions couple in a flavour universal way
to the $\mathcal{G}_\mu^A$:
\begin{align}
{J}^{\mu\,a}_U  ~=~  &   \bar Q^{u\,a}_i   \gamma_{\mu}  L^\nu_i  +   \bar Q^{d\,a}_i   \gamma_{\mu}  L^e_i\,,  \\
J_{G}^{\mu\,\hat{a}} ~= ~  &    \,  \bar Q_i^u   T^{\hat{a}} \gamma_{\mu}  Q_i^u +  \bar Q_i^d   T^{\hat{a}} \gamma_{\mu}  Q_i^d\,,   \\
{J}^\mu_{X} ~= ~&  \bar Q_i^u \gamma_{\mu}  Q_i^u +  \bar Q_i^d \gamma_{\mu}  Q_i^d  {- 3}  \bar L_i^\nu \,   \gamma_{\mu} L_i^\nu {- 3}  \bar L_i^e \,   \gamma_{\mu} L_i^e\,.
\label{currentsX}
\end{align}
 \item
 The allowed Yukawa couplings are between vector-like and chiral fermions,
\begin{equation}
\mathcal{L}_Y = H\bar{Q}^u_LY_u u_R + H \bar{L}^{\nu}_LY_\nu \nu_R+ H^*\bar{Q}^d_LY_d d_R+ H^*\bar{L}^e_LY_e e_R,
\label{Yuk}
\end{equation}
and they are assumed to respect a product $U(2)_{F_u+u+\nu}\times U(2)_{F_d+d+e}$  of diagonal symmetries.
The minimal breaking of the overall $U(2)_q\times U(2)_l\times U(2)_{F_u+u+\nu}\times U(2)_{F_d+d+e}$ flavour symmetry is 
controlled by 
\begin{equation}
\mathcal{L}_{\rm mix}=\bar{q}_L\hat{m}_u Q^u_R + \bar{q}_L\hat{m}_d Q^d_R + \bar{l}_L\hat{m}_e L^e_R + \bar{l}_L\hat{m}_\nu L^\nu_R \,,
\label{only_break}
\end{equation}
with the mixing matrices $\hat{m}_{u,d,e,\nu}$ having the same flavour structure as the Yukawa couplings in Eq.~(\ref{Yuka}).
\end{itemize}
\section{Couplings of the massless fermions}
In this Section, we derive the couplings of the fermions which are massless before ElectroWeak symmetry breaking. Needless to say, their couplings to the SM gauge bosons are fixed by gauge invariance. We thus need to derive the couplings to the  $\mathcal{G}_\mu^A$ and the Yukawa couplings, which are not flavour generic.
\subsection{Model 1}
Without loss of generality, we can take the spurion doublet $V_q$ oriented in the direction of $q_{2}$ with a single entry $\epsilon_q = O(|V_{cb}|)$. This way, the mixing part in Eq.~(\ref{yandm}) involves only the 2--3 sector and can be put in the form
\begin{equation}
\mathcal{L}_{\rm mix}=\bar{q}_{L}  U_qm_qW_q^+Q_R\,, \qquad 
q_L= \left(\begin{array}{c}
q_{2}\\
q_{3}
\end{array}\right)\,, \quad
Q_R= \left(\begin{array}{c}
Q_{1_R}\\
Q_{2_R}
\end{array}\right)\,, 
\label{Lmix1}
\end{equation}
where
\begin{equation}
m_q = M  \left(\begin{array}{cc}
O(\epsilon_q)&0\\
0& O(1)
\end{array}\right),
\end{equation}
and $U_q$ and $W_q$ are $2\times 2$ unitary matrices with off-diagonal elements of order $\epsilon_q$ and $1$, respectively.
Note that, differently from the previous section, here $q_L$ denotes a two-component vector. We will go back to the three-family notation at the end of this section.

After inserting Eq.~(\ref{Lmix1}) into (\ref{yandm}), the complete $\mathcal{L}_m$ leads to two massive states
\begin{equation}
Q^h_L= s_q U_q^+q_L +c_q W_q^+ Q_L\,, \quad s_q= \frac{m_q}{\sqrt{m_q^2 + M^2}}\,,
\end{equation}
and two massless orthogonal combinations
\begin{equation}
q^l_L= c_q U_q^+q_L - s_q W_q^+ Q_L\,,
\end{equation}
which acquire a mass only via the Yukawa interaction.
By inverting these equations, the components of the (interaction) fields $Q_L$ and $q_L$ involving 
the light states  are 
\begin{equation}
\left.  Q_L \right|_{\rm light} =-W_qs_q q^l_L\,, \qquad   \left. q_L \right|_{\rm light} =U_q c_q q^l_L\,.
 \label{light}
\end{equation}
Analogous expressions hold for the leptons.

The insertion of  $Q_L |_{\rm light}$ and $L_L |_{\rm light}$ into the currents in Eq.~(\ref{SU4currents}) yields the couplings of the light states to the $SU(4)$ vectors. To simplify the notation we remove the suffix $l$ from the fields.
For the coloron and the $B$--$L$ current there are no flavor-changing couplings:
\bea
J_\mu^{G\,\hat{a}} &=& \bar{q}_Ls_q^2\gamma_\mu T^{\hat{a}} q_L\,,  \label{JmuG} \\
J_\mu^{X} &=& \bar{q}_Ls_q^2\gamma_\mu q_L -3 \bar{l}_Ls_l^2\gamma_\mu l_L\,.
\label{JmuX}
\eea
The leptoquark current instead has the form:
\begin{equation}
J_\mu^{U\,a} = \bar{q}_L^a s_q Ws_l\gamma_\mu  l_L\,,  \quad\quad W=W_q^+W_l\,.
\label{JmuU}
\end{equation}
Since  $q_{1}$ does not  enter $\mathcal{L}_{\rm mix}$, we have  $q_{1}=q_{1}^l$, hence in these expressions for the $SU(4)$ currents $q_L$ and $l_L$ can be thought to include all the three generations with $s_{q_1}=s_{l_1}=0$. 
Note that here the $q_L$ field does not correspond either to the up- or to the down-quark mass eigenstates, since we have not fully diagonalised the Yukawa couplings yet; similar considerations hold for $l_L$.
It is readily seen that the transformation $q_L\rightarrow U_q c_q q^l$ in the 2--3 sector does not alter the structure of the Yukawa couplings in (\ref{Yuka}). These are diagonalised by a proper unitary matrix on the left side only\footnote{~A unitary transformation in the $(1,2)$ right-handed sector is unphysical since it can be taken away by a proper redefinition of $u_R, d_R, e_R$.}.
\begin{equation}
\hat{y}_d = U_d  y_d^{\rm diag}\,, \qquad \hat{y}_u = U_u  y_u^{\rm diag}\,,
\qquad \hat{y}_e = U_e  y_e^{\rm diag}\,,
\end{equation}
where $U_u =  U_d V_{\rm CKM}^\dagger$. As shown in Ref.~\cite{Fuentes-Martin:2019mun}, 
$U_d$ and $U_e$ have the following parametric form:
\be
U_d \approx
\begin{pmatrix}
 c_d   &  -s_d\,e^{i\alpha_d}  & 0  \\
 s_d\,e^{-i\alpha_d}   &  c_d &  s_b   \\
-s_d\,s_b\,e^{-i(\alpha_d+\phi_q)}   & -c_d\,s_b\, e^{-i\phi_q} & e^{-i\phi_q}
\end{pmatrix}\,,
 \qquad U_e \approx 
\begin{pmatrix}
c_e       & -s_e          & 0  \\
s_e &  c_e           & s_\tau \\
-s_e s_\tau   &  -c_es_\tau & 1
\end{pmatrix}\,. 
\label{eq:Ud}
\ee
These expressions are obtained expanding up to first non-trivial terms in the small mixing 
parameters $s_{d,b,e,\tau}$ with $s_i=\sin\theta_i$,  
$c_i=\cos\theta_i$. The known structure of the CKM matrix implies 
 that $s_d$ and $\alpha_d$ are not free but are constrained by 
 \be
\frac{s_d}{c_d}=\left| \frac{V_{td}}{V_{ts}}\right| \,,  \qquad \alpha_d=\arg(V_{td}^*/V_{ts}^*)\,.
\ee
On the other hand, the  $2$--$3$ mixing angles, which 
are related to the Yukawa parameters by $s_b/c_b=|x_b|\,|V_q|$ and $s_\tau/c_\tau=|x_\tau|\,|V_\ell|$, 
cannot be expressed in terms of SM observables. Unconstrained are also $s_e$ and 
$\phi_q$ (that becomes unphysical in the limit $s_b\to 0$). 

The $3\times3$ unitary matrices $U_{d,u,e}$ appear in the currents~(\ref{JmuG})--(\ref{JmuU}) once these are 
written in terms of mass eigenstates.  For instance, the coloron current involving 
down-type quarks takes the form
\be
\left. J_\mu^{G\,\hat{a}} \right|_{\rm down} = \bar{d}_L\, U^\dagger_d s_q^2\gamma_\mu T^{\hat{a}} U_d\,  q_L\,. 
\ee
%
\subsection{Models 2 and 3}
We can effectively discuss these two models together by assuming two families 
of vector-like fermions and later treating $F_1$ as a null entry in Model~2.
The part of $\mathcal{L}_m$ that contains only $SU(4)$ 4-plets $( f_3, F_1, F_2)$  
can be diagonalised by the unitary transformation 
\begin{equation}
 \left(\begin{matrix} q_3 \\Q_{1_L}\\Q_{2_L} \end{matrix} \right) = W_q  \left(\begin{matrix} q_3^l \\Q_{1_L}^h\\Q_{2_L}^h \end{matrix} \right)\,, 
\end{equation}
and similarly in the $2\times 2$ $Q_R$ sector. We can proceed analogously in the lepton sector. 

In this new basis, the coloron and the $X$ currents remain universal and diagonal, whereas the leptoquark current is modified by a unitary matrix $W_q W_l^+\equiv W$.
Also, mass mixing occurs only in the $2\times 2$ sector:
\begin{equation}
\mathcal{L}_m = M\bar{q}_{i}\tilde{\Delta}_{ij}Q_{j_R}^h + \bar{Q}_{j_L}^h M_{Qj} Q_{j_R}^h \,, 
\end{equation}
where
\be
\left. \tilde{\Delta}  \right|_{\rm Model\ 2}= O(\epsilon_q) \times \left(\begin{array}{cc} 0 &0\\
0 &   1 \end{array}\right)\,, \qquad 
\left. \tilde{\Delta}  \right|_{\rm Model\ 3}= O(\epsilon_q) \times \left(\begin{array}{cc} O(\lambda^2) & O(\lambda)\\
O(\lambda) & 1\end{array}\right)\,, \qquad 
\ee
with $\lambda=|V_{us}|$. Diagonalising $\mathcal{L}_m$ we obtain the light states $q^l_L\approx q_L-\hat{\Delta}_q Q^h_L$ and the heavy orthogonal combinations $\hat{Q}^h_L\approx Q^h_L+\hat{\Delta}_q^\dagger q_L$, with  $\hat{\Delta}_q=\tilde{\Delta}M/M_Q$. Up to irrelevant unitary transformations from the right, we can write
\be
\left.   \hat{\Delta}_q  \right|_{\rm Model\ 2}=  \left(\begin{array}{cc} 0 &0\\
0 &   s_{q_2} \end{array}\right)\,, \qquad 
\left.   \hat{\Delta}_q  \right|_{\rm Model\ 3}= 
 \hat U
 \left(\begin{array}{cc}
s_{q_1}&0\\
0&s_{q_2}
\end{array}\right)\,,
\ee
where  $\hat U$ is a (complex) unitary $2\times 2$ matrix with off-diagonal entries of $O(\lambda)$.

As in the previous case, one obtains the final form of the
 $SU(4)$ currents by expressing the interaction fields  
in terms of their light components. In Model 2, this leads exactly to the same currents as in Model 1 with $s_{q_3}=s_{l_3}=1$ (and $s_{q_1}=s_{l_1}=0)$. At the same time, the structure of the Yukawa couplings remains unchanged. 

In Model 3 there are three non vanishing hierarchical angles $s_{q_1}< s_{q_2}< s_{q_3}$ and, most important, a unitary transformation in  the $1$--$2$ sector.  The latter can be moved to the Yukawa sector,
recovering a flavour-diagonal structure for the $SU(4)$ currents as in Eqs.~(\ref{JmuG})--(\ref{JmuU}),
but altering the structure of the Yukawa couplings. This results in a difference in the Yukawa 
diagonalization matrices that is particularly relevant in the 1--2 quark sector.
In the basis where the  $SU(4)$ currents have the form in Eqs.~(\ref{JmuG})--(\ref{JmuU}),
\begin{equation}
\hat{y}_d = \hat{U}_d\  y_d^{\rm diag}, \qquad \hat{U}_d =
\left(\begin{array}{c:c}
 \hat U^\dagger  & 0 \\\hdashline
 0 & 1
\end{array}\right) 
\times 
 U_d\,.
 \end{equation}
The matrix $\hat{U}_d$ has the same parametric form of $U_d$ in Eq.~(\ref{eq:Ud}),
with $O(\lambda)$ 1--2 off-diagonal entries, and 2--3 and 1--3 entries of order $O(s_b)$ and 
$O(\lambda s_b)$, respectively. Unlike in Model 2, however,  the entries in the $2\times 2$ 
(light-family) block are now free parameters,  not constrained by the CKM matrix elements. In particular, 
we can reach the limit of real mixing in the light-family sector (i.e.~the limit where the 
$2\times 2$ light-family block of $\hat{U}_d$ is a real orthogonal matrix) 
that, as we shall see and as recently pointed out in Ref.~\cite{Crosas:2022quq}, is phenomenologically favoured.

It is worth stressing that real mixing in the light-family sector can be obtained if all the $U(2)^n_f$ breaking 
bi-doublets in the quark sector ($\Delta_{u,d}$ and $\Delta^q$) are CP conserving.
In this case, the phase of the CKM matrix originates only from the $V_q$ spurion, which in the basis where 
all the bi-doublets are real has a non-trivial orientation in $U(2)_q$ space.\footnote{~In this limit, the 
phase $\alpha_d$ in the expression of $U_d$ in Eq.~(\ref{eq:Ud}) appears as a consequence of having 
chosen a $U(2)_q$  basis where  $V_q$ is oriented in the direction of the second family.}
\subsection{Model 4}
Let us define the 9-component vector made of the left-handed fields
\begin{equation}
Q=
  \left(\begin{matrix} q \\ Q^u \\  Q^d  \end{matrix} \right),
\end{equation}
 as indicated. The mass matrix in the quark sector before ElectroWeak symmetry breaking, $\mathcal{L}_m^q$, is diagonalised from the left by a unitary transformation of the form $Q_m= WQ$, where
\begin{equation}
Q_m=
  \left(\begin{matrix} q^l \\ Q^u_m \\  Q^d_m  \end{matrix} \right)\,,
  \label{Qm}
\end{equation}
and $q^l$ are the three massless states. 

As in the previous cases, to determine the  
interactions of the light states both with the Higgs and with the $SU(4)$ vectors, 
what counts are the light components in $Q^{u,d}$:
\begin{equation}
\left.   Q^u\right|_{\rm light}  = w_uq^l\,, \qquad    \left.   Q^d \right|_{\rm light} =w_dq^l\,,
\end{equation}
where $w_{u,d}$ are $3\times 3$ matrices. With similar considerations in the lepton sector, one obtains the Yukawa couplings
\begin{equation}
\hat{y}_u=w_u^+Y_u\,,\qquad
\hat{y}_d=w_d^+Y_d\,,\qquad
\hat{y}_\nu=w_\nu^+Y_\nu\,,\qquad
\hat{y}_e=w_e^+Y_e\,.
\end{equation}
Since $Y_{u,d,\nu,e}$ are diagonal matrices, the $SU(4)$ currents reads ($q^l, l^l\rightarrow q, l$)
\bea
J_\mu^{U\,a} &=& \bar{q}^a \left( \hat{y}_u\frac{1}{Y_uY_\nu}\hat{y}_\nu^+ + \hat{y}_d\frac{1}{Y_dY_e}\hat{y}_e^+ \right)\gamma_\mu l\,,
\label{J^U}  \\
J_\mu^{G\,\hat{a}}  &=& \bar{q} \left(\hat{y}_u\frac{1}{Y_u^2}\hat{y}_u^+ + \hat{y}_d\frac{1}{Y_d^2}\hat{y}_d^+ \right)\gamma_\mu T^{\hat{a}} q \,,
\label{J^G} \\
J_\mu^X &=& \bar{q} \left(v\hat{y}_u\frac{1}{Y_u^2}\hat{y}_u^+ + \hat{y}_d\frac{1}{Y_d^2}\hat{y}_d^+)q
-3 \bar{l}(\hat{y}_\nu\frac{1}{Y_\nu^2}\hat{y}_\nu^+ + \hat{y}_e\frac{1}{Y_e^2}\hat{y}_e^+ \right)l\,.
\label{J^X}
\eea
At the same time, as implied by the spurion transformation properties under $U(2)^n_f$, the Yukawa couplings assume the usual structure
\begin{equation}
\hat{y}_u\approx U_u y_uW_u^+\,,\quad
\hat{y}_d=  U_d y_dW_d^+\,,\quad
\hat{y}_\nu\approx U_\nu y_\nu W_\nu^+\,,\quad
\hat{y}_e=  U_e y_eW_e^+\,,
\label{ys}
\end{equation}
where $W_{u,d,\nu,e}$ are unitary matrices in the $2\times2$ light-family sector only. 
Defining
\begin{equation}
U_e^\dagger  U_\nu \equiv E\,,
\end{equation}
and recalling that  $U_u^\dagger U_d = V \equiv V_{\rm CKM}$, the $SU(4)$ currents, 
written in terms of the light mass eigenstates, are given by
\bea
J_\mu^{G\,\hat{a}} &=&  \bar{u}_L (z_u^2 +  Vz_d^2 V^\dagger)\gamma_\mu T^{\hat{a}} u_L +
\bar{d}_L  (z_d^2 +  V^\dagger z_u^2 V)\gamma_\mu T^{\hat{a}} d_L\,, \\
J_\mu^{U a} &=&  \bar{u}^a_L (z_uW_{u\nu}z_\nu E^\dagger +  V z_dW_{de}z_e )\gamma_\mu \nu_L +
\bar{d}^a_L  (V^\dagger z_uW_{u\nu}z_\nu E^\dagger +  z_dW_{de}z_e )\gamma_\mu e_L\,, \\
J_\mu^X &=&   -3\bar{e}_L (z_e^2 +  Ez_\nu^2 E^\dagger)\gamma_\mu e_L -
3\bar{\nu}_L  (z_e^2 +  Ez_\nu^2 E^\dagger)\gamma_\mu \nu_L+ \\
&& \bar{u}_L (z_u^2 +  Vz_d^2 V^\dagger)\gamma_\mu u_L +
\bar{d}_L  (z_d^2 +  V^\dagger z_u^2 V)\gamma_\mu d_L\,,\nonumber
\eea
where 
\begin{equation}
z_{u,d,\nu,e}=  \left. \frac{y^{\rm diag}}{Y} \right|_{u,d,\nu,e}\,,
\qquad W_{u\nu}= W_u^\dagger W_\nu\,,
\qquad W_{de}= W_d^\dagger W_e\,.
\end{equation}
Note that, due to $U(2)^n_f$ invariance of the Yukawa matrices in Eq.~(\ref{Yuk}), we have:
\begin{equation}
\frac{z_{u1}}{z_{u2}}= \frac{m_u}{m_c}\,, \quad
\frac{z_{d1}}{z_{d2}}= \frac{m_d}{m_s}\,, \quad
\frac{z_{e1}}{z_{e2}}= \frac{m_e}{m_\mu}\,.
\end{equation}
Note also the asymmetry between quarks and leptons in the above equations, given that the $\nu_L$ are defined as current eigenstates.
\section{Phenomenological analysis}
\begin{table}[t]
\centering
\renewcommand{\arraystretch}{1.2} 
\scalebox{0.90}{
\begin{tabular}{cccccc}
\hline
Class & { Observable} & { Experiment/constraint}  & Correlation &  { SM prediction}  & { Theory expr.}  \\
\hline 
\hline
 & $ C_{9, \mathrm{NP}}^\mu = - C_{10,\mathrm{NP}}^\mu$ & $-0.39 \pm 0.07$ \cite{Altmannshofer:2021qrr} & $-$ & $0$ & \eqref{eq:C9}  \\
I & $  R_D $  &  $  0.340 \pm 0.030 $   \cite{HFLAV:2022pwe}  &  $ \rho = -0.38$ &   $0.298 \pm 0.003$ \cite{HFLAV:2022pwe} & \eqref{eq:RDRDs}   \\
&  $R_{D^{\ast}}$ & $0.295 \pm 0.014$   \cite{HFLAV:2022pwe} &&   $0.252 \pm 0.005$  \cite{HFLAV:2022pwe}  &  \eqref{eq:RDRDs}  \\ \hline
II & $  (g_{\tau}/g_{e,\mu})  $ & $1.0012 \pm 0.0012$ \cite{HFLAV:2022pwe} & $-$ & $1$ & \eqref{eq:tauLFUV}  \\  \hline
\multirow{2}{*}{III} & $  \tau \to 3 \mu $ & $< 2.1 \times 10^{-8}$ \cite{ParticleDataGroup:2020ssz} & $-$ & $0$ & \eqref{eq:tau23mu}    \\ 
& $ K_L \to \mu^{\pm} e^{\mp}  $ & $<4.7 \times 10^{-12}$ \cite{ParticleDataGroup:2020ssz}   & $-$ & $0$ &      \\  \hline
 & $\delta (\Delta m_{B_s})$  &  $ 0.0 \pm0.1$  [*] & $-$ & $0$ & \eqref{eq:DmBs} \\
IV & ${\rm Im}(\mathcal{C}_{uc}^{\rm NP})~[{\rm GeV}^{-2}]$ &  $(-0.03 \pm 0.46)\times10^{-14}$  \cite{FlavConstraints,UTfit:2007eik} &$-$ & $0$ &  \eqref{eq:DF2}   \\
& ${\rm Im}(\mathcal{C}_{ds}^{\rm NP})~[{\rm GeV}^{-2}]$ &  $ (0.06 \pm 0.09)\times10^{-14}$    \cite{FlavConstraints,UTfit:2007eik} &$-$& $0$ &  \eqref{eq:DF2}  \\
\hline
\end{tabular}
}
\caption{\label{tab:obs} Relevant low-energy observables.  Upper bounds on branching ratios correspond to 90\% CL. The entry marked with a [*] denotes our constraint imposed on the magnitude of $\Delta m_{B_s}$. }
\end{table} 
We now proceed with a phenomenological analysis of the most relevant observables for the different models. 
We limit our analysis to observables that are either mediated at the tree-level by the new massive vectors, 
or are loop-induced but dominated by logs insensitive to the UV completion. The overall size of the 
non-standard contributions is controlled by the effective coupling $C_U$ or, equivalently, by the effective scale 
$m_{\rm{eff}} = m_U/g_U$, defined as
\begin{equation}
C_U = \frac{g_U^2}{m_U^2}  \frac{m_W^2}{g^2}  = \frac{v^2}{4 m_{\rm{eff}}^2}\,. 
\end{equation}
To unify the description of the relevant observables, 
we introduce the effective couplings $\beta^V_{ij}$ parameterizing the currents of the heavy vectors 
to the light mass-eigenstates:
\bea
 J_\mu^{U\, a} = \beta^U_{i \alpha}\,   \bar q^{i\, a}_{L}   \gamma_{\mu}   \ell^\alpha_{L}~,  
\qquad  J^{G\, \hat{a}}_{\mu} = \beta^G_{ij}\,   \bar q^i_{L}   \gamma_{\mu}  T^{\hat{a}}  q^j_{L}~, \qquad 
 J^{X}_\mu =  \beta^X_{ij}\,   \bar q^i_{L}  \gamma_{\mu}  q^j_{L}~  + 
			  \beta^X_{\alpha \beta}\,  \bar \ell^\alpha_{L}\,   \gamma_{\mu}  \ell^\beta_{L}\, .
\eea
The expressions of the $\beta^V_{ij}$ in the different models can be readily derived from the previous section.
Simplified expressions for the relevant observables in terms of the $\beta^V_{ij}$ are reported in Appendix \ref{app:obs}.\footnote{~To highlight the origin of the specific mediator contributing to each observable, the formulae 
 in Appendix \ref{app:obs} are expressed in terms of the three overall couplings $C_{X}$, $C_{G^\prime}$, $C_U$, 
 which are assumed to be identical in the fit.}
The observables playing a relevant role in constraining the model parameters, given the current experimental
data, are collected in Table \ref{tab:obs}. We can divide them into four groups:
\begin{itemize}
\item[I]~LFU anomalies. This group is the only one providing indications for non-vanishing  $\beta^V_{ij}$.
In the case of $b \to s \ell^+\ell^-$ observables, we use as inputs the best-fit value of the modified Wilson coefficients 
$C_{9,10}$ extracted from a global fit to $b \to s \ell^+\ell^-$ data in Ref.~\cite{Altmannshofer:2021qrr},
 assuming a pure left-handed structure for the non-standard contribution ($C_{9, \mathrm{NP}}^\mu = - C_{10,\mathrm{NP}}^\mu$ ).
In the case of $b \to c \tau \nu$ observables, we use the HFLAV averages for the experimental results and for the
theoretical predictions of $R_{D^{(*)}}$~\cite{HFLAV:2022pwe}.
\item[II]~Tests of universality in $\tau$ decays. These are expressed via the effective coupling ratio 
$(g_{\tau}/g_{e,\mu})$~\cite{HFLAV:2022pwe}. Despite not receiving tree-level contributions in our setup, 
a largely model-independent mo\-di\-fication of $(g_{\tau}/g_{e,\mu})$ is generated at the one-loop level. As pointed out first in Ref.~\cite{Feruglio:2016gvd}, this effect provides 
a significant constraint to any model addressing the $R_{D^{(*)}}$ anomalies.
\item[III] ~Lepton Flavour Violating (LFV) rates. The most relevant observables to constrain $\tau$--$\mu$
and $\mu$--$e$ couplings in our models are the bounds on $\mathcal{B}( \tau \to 3 \mu )$ and  $ \mathcal{B}(K_L \to \mu e)$.
\item[IV] $\Delta F=2 $ observables. In this case, the most relevant constraints are derived from 
$\Delta m_{B_s}$, for which we assume a reference $10\%$ Gaussian error over its SM value, and 
CP violation in neutral $K$- and $D$-meson mixing. In the latter case, we use as inputs the constraints on 
the imaginary parts of four-quark left-handed Wilson coefficients determined in Refs.~\cite{FlavConstraints,UTfit:2007eik}, in the standard convention for the phases of the CKM matrix elements.
\item In all models it is easy to accommodate the present central values of the $b \to s \ell^+\ell^-$ anomalies, and $\tau \to 3 \mu$ does not pose a serious constraint
\end{itemize}
 \begin{figure}[t]
\centering  
\includegraphics[scale = 0.38]{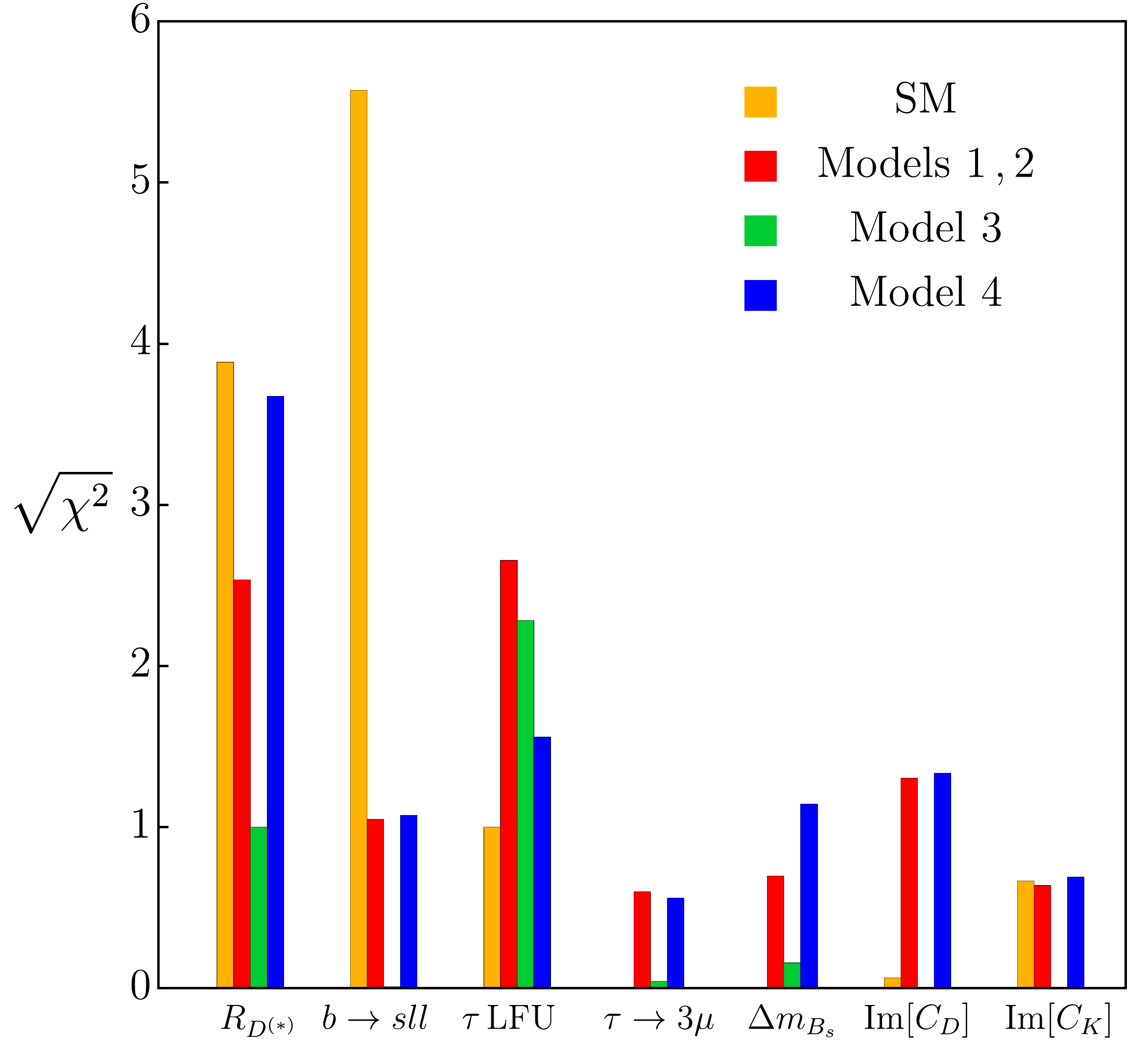}
\caption{Contribution to $\sqrt{\chi^2}$ coming from each observable for models 1-4.}
\label{fig:Chi2_allcases}
\end{figure}
We distinguish three different non-SM fits, given that Models 1 and 2 are described by an equivalent set of effective parameters. The contribution to $(\chi^2_{\mathrm{min}})^{1/2}$ (or the pull of the fit)  from the various observables in the three different cases, vs.~the SM one, is reported in Figure \ref{fig:Chi2_allcases}.   The best fit points for the model parameters are reported in Table \ref{tab:fit}.

A few details about the fit procedure are in order. Since $s_e$ is constrained only by $K_L\to \mu e$, in all cases 
we set $s_e = 0$ and treat this parameter (as well as $K_L\to \mu e$)  separately.  In models 1,2,3 we assume $s_{l_1} = s_{q_1} = 0$ and $s_{l_3} = s_{q_3} = 1$. 
We further implement the requirement that $U(2)^n_f$ breaking should not exceed its natural size dictated by the 
structure of the SM Yukawa couplings,  adding a smooth gaussian contribution with $\sigma = 0.05$ to the $\chi^2$ for $\abs{s_{\tau, b, q_2,l_2}} > 0.1$ and $\abs{\theta_d}> 0.3$. Additionally, we impose a smooth gaussian contribution for $\abs{\theta_{\chi_2}} > 0.5$.  For Model 4, on top of adding a smooth gaussian contribution with $\sigma = 0.05$ to the $\chi^2$ for $\abs{s_{\tau}} > 0.1$,  we assume  the following strict boundaries $\abs{z_{u_2,d_2,e_2,\nu_2}}< 0.2$,  
 $0.5< \abs{z_{d_3,u_3, e_3,\nu_3}}< 1$,  and $\abs{z_{\nu_1}} < 0.01$, again dictated by  a natural flavour symmetry breaking  
structure.
\begin{table}[p]
\centering
\renewcommand{\arraystretch}{1.1} 
\scalebox{0.95}{
\begin{tabular}{ccccc} 
\hline
case & $\chi^2_{\rm{min}}$ & model parameters & best fit point  & $1 \sigma$  \\
\hline
 \hline
 \multirow{8}{68 pt}{\centering{  \textbf{Model 1,2}     $s_d$, $\alpha_d$ fixed \\ $ \theta_{\chi_2}$ free  }}  &  \multirow{7}{*}{$20.9 (17.5)$}   
 &  $m_{\rm eff}$ [TeV]  &   $0.76$ & $ [0.59,0.80] $  \\
 && $s_{q_2}$ & $0.02$ &\\
 && $s_{l_2}$ & $0.18$ &\\
 && $s_\tau$ & $0.12$ &\\
 && $s_b$ & $0.004$ &\\
  && $\theta_{\chi_2}$ & $0.50$ &\\
 && $\alpha_\chi$ & $0.66$ &\\
 \hline
  \multirow{8}{68 pt}{\centering{ \,\, \textbf{Model 3}  \,\,   $s_d$, $\alpha_d$ free \\  $ \theta_{\chi_2} $ free  }}  &  \multirow{10}{*}{$6.8 (6.2)$}   
 &  $m_{\rm eff}$ [TeV]  &   $0.67$ & $ [0.66,0.92] $  \\
 && $s_{q_2}$ & $0.13$ &\\
 && $s_{l_2}$ & $0.10$ &\\
 && $s_\tau$ & $0.10$ &\\
 && $s_b$ & $-0.002$ &\\
 && $\theta_{\chi_2}$ & $0.51$ &\\
 && $\alpha_\chi$ & $0$ &\\
  && $\theta_d$ & $-0.24$ &\\
 && $\alpha_d$ & $0.00005$ &\\
\hline
\multirow{12}{65 pt}{\centering{ \,\, \textbf{Model 4}  \,\,   minimal model }}  &  \multirow{11}{*}{$23.7 (20.9)$}   
  &  $m_{\rm eff}$ [TeV]  &   $1.31$ & $ [0.93,1.35] $  \\
 && $z_{u_2}$ & $0.02$ & \\
 && $z_{u_3}$ & $0.5$ &\\
  && $z_{d_2}$ & $-0.07$ &\\
 && $z_{d_3}$ & $-0.5$ &\\
   && $z_{e_2}$ & $0.2$ &\\
 && $z_{e_3}$ & $-1$ &\\
  && $z_{\nu_1}$ & $-0.01 $ &\\
    && $z_{\nu_2}$ & $-0.2$ &\\
      && $z_{\nu_3}$ & $1$ &\\
 && $s_\tau$ & $0.17$ & \\
\hline
\end{tabular}}
\caption{Fit results for the different Models. In the $\chi^2_{\mathrm{min}}$ entry, the number in parenthesis is the contribution to $\chi^2_{\mathrm{min}}$ that comes solely from the observables.  For reference, in the SM $\chi^2_{\rm{min, SM}} = 47.6$. }
\label{tab:fit}
\end{table}
\begin{figure}[t]
 \centering
\includegraphics[scale = 0.38]{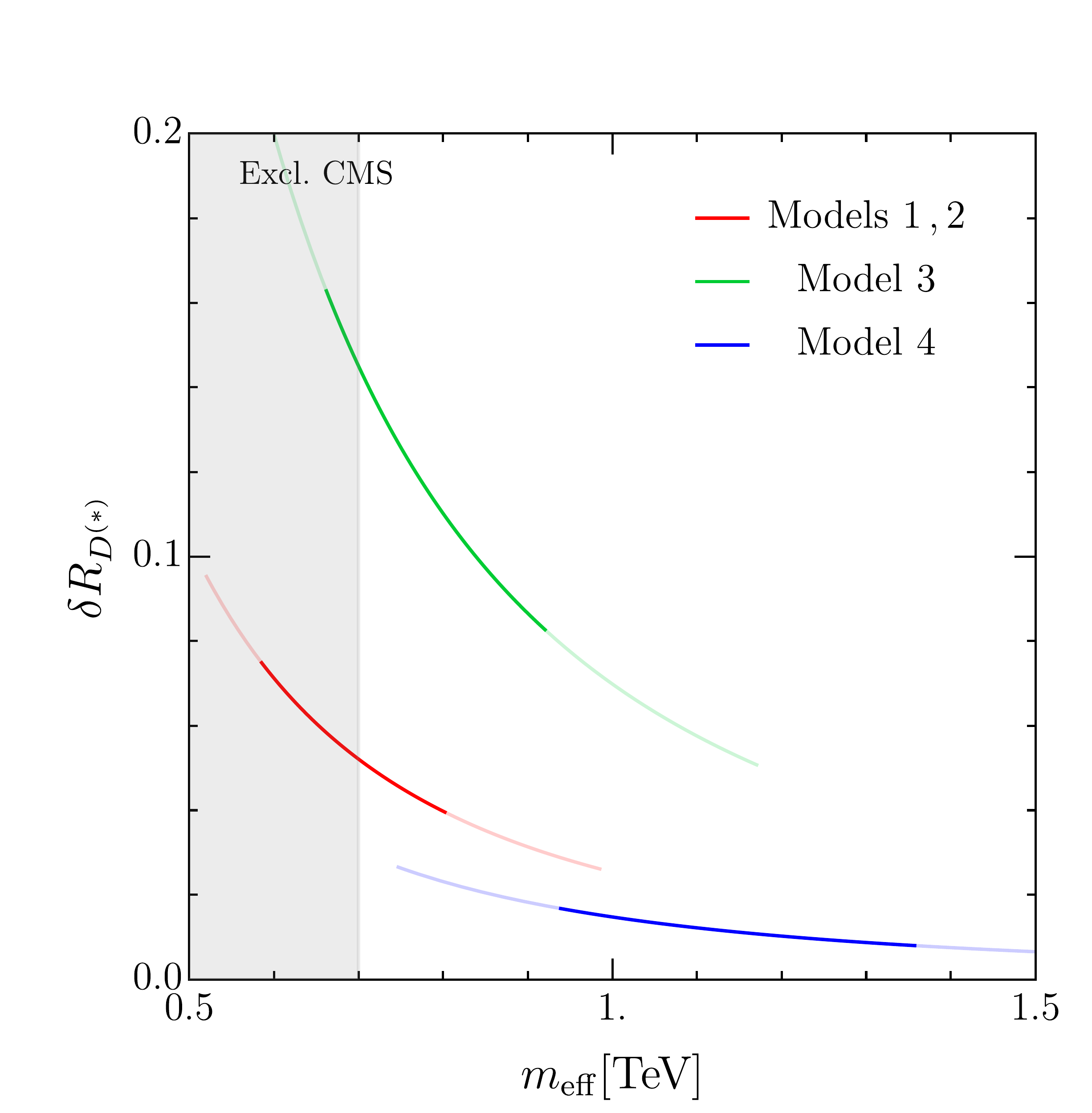}
\caption{$\delta R_{D^{(\ast)}}$ as a function of $m_{\rm eff}$ varying $m_{\rm eff}$ in the $1 \sigma$ (dark)  and $2 \sigma$ (light) region preferred by the fit. The remaining parameters entering $\delta R_{D^{(\ast)}}$ are fixed to the best-fit point.   The grey band indicates the CMS exclusion at 700 GeV~\cite{CMS:2022a,CMS:2022b,CMS:2022c}. }
\label{fig:RDmeff_allcases}
\end{figure} 
\section{Discussion}
From the analysis of the fit results, illustrated in Figures~\ref{fig:Chi2_allcases}  and Table~\ref{tab:fit}, we can deduce the following conclusions:

\begin{itemize}
\item In all models a strong lower bound on $m_{\rm eff}$ is set by LFU in $\tau$ decays. This limits the contribution to  $\delta R_{D^{(\ast)}}$  from 
the pure third-generation semileptonic operator (in the down-quark mass basis)  generated by the leptoquark exchange,
i.e.~the amplitude proportional to  $C_U V_{cb} \beta^{U}_{b \tau}  \beta^{U \ast}_{b \tau}$ in Eq.~(\ref{eq:RDRDs}).
\item  In models 1, 2 and 3, if $\theta_{\chi_2}  = O(1)$ and $s_{q_2}\not=0$, a sizeable effective coupling $\beta^{U}_{s \tau} \approx s_{\chi_2} s_{q_2}$
 is generated. This can  in turn yield an additional contribution to 
$\delta R_{D^{(\ast)}}$ proportional to $C_U V_{cs} \beta^{U}_{s \tau}  \beta^{U \ast}_{b \tau}$.   
In Model 4, since there is no freedom in the flavour-changing interactions, we do not have such a possibility. Indeed, in this case it is not possible to generate a sizeable contribution to $R_{D^{(\ast)}}$. Additional constraints from $\Delta F=2$ observables drive the best-fit point for this model toward higher $m_{\rm eff}$ values. 
\item In models 1, 2 and 3, in order to reach the present central value of $\delta  R_{D^{(\ast)}}$, we need not only $\theta_{\chi_2}  = O(1)$ 
but also $s_{q_2} = O(0.1)$. The latter condition is phenomenologically viable only if we can adjust the parameters 
in $U_d$ (or $\hat U_d$) to minimize the $\Delta F=2$ amplitudes generated by the 
(tree-level) coloron exchange.  While all models can satisfy the $B_s$ mixing bound
in the $s_b\to0$ limit, only in Model 3 is possible to adjust also $s_d$ and $\alpha_d$ in order to satisfy both $K-\bar{K}$ and $D - \bar{D}$ mixing bounds.
As anticipated, this happens in the case of a real $2\times 2$ light block in $\hat U_d$.
\item Also the constraint from $K_L \to \mu e$ is easily evaded. In particular, for the best fit point of Model~3, the $K_L \to \mu e$ bound implies 
the (weak) condition $|s_e|<0.6$.
\end{itemize}

\begin{figure}[t]
\centering  
\includegraphics[scale = 0.35]{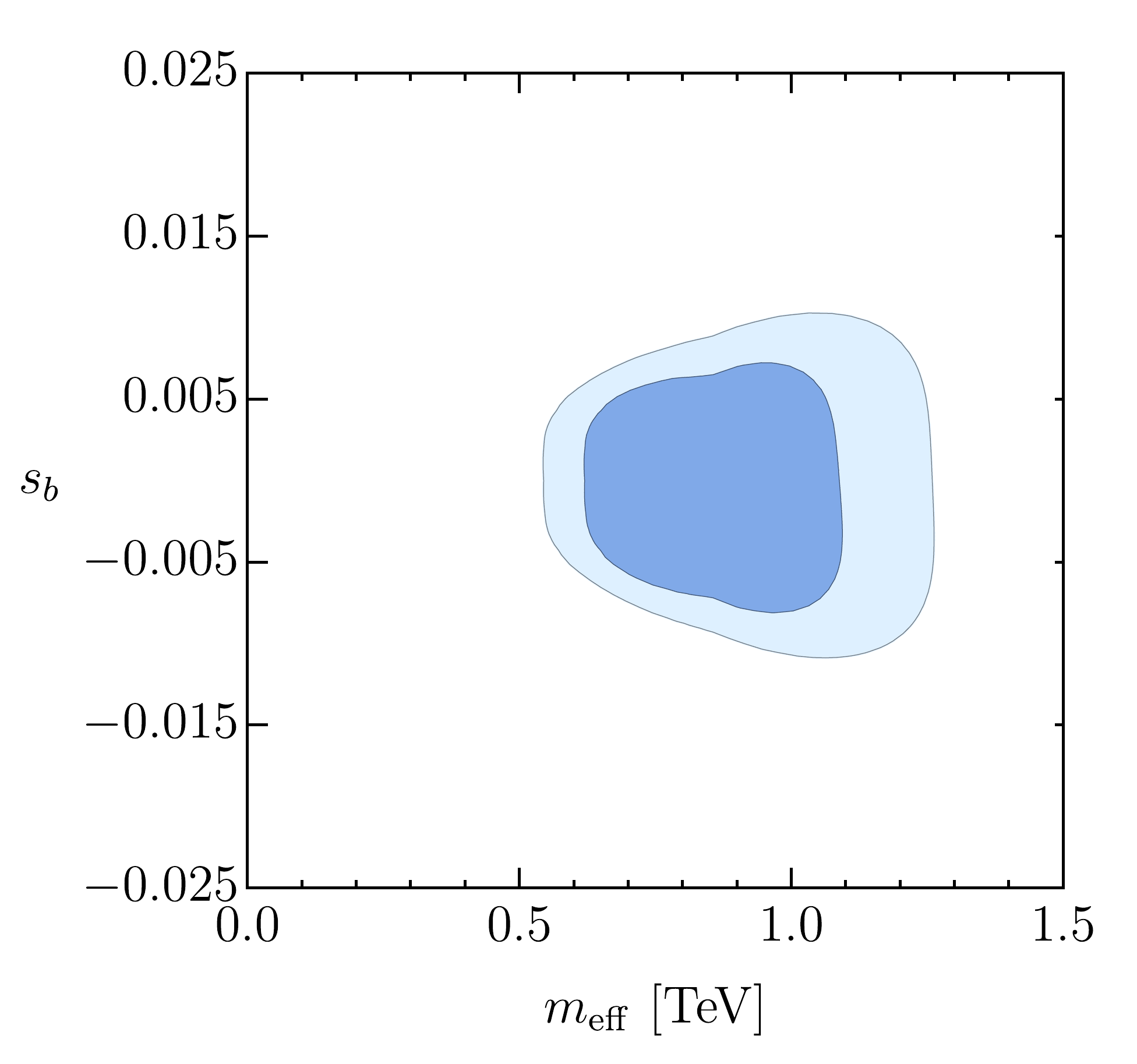}
\includegraphics[scale = 0.35]{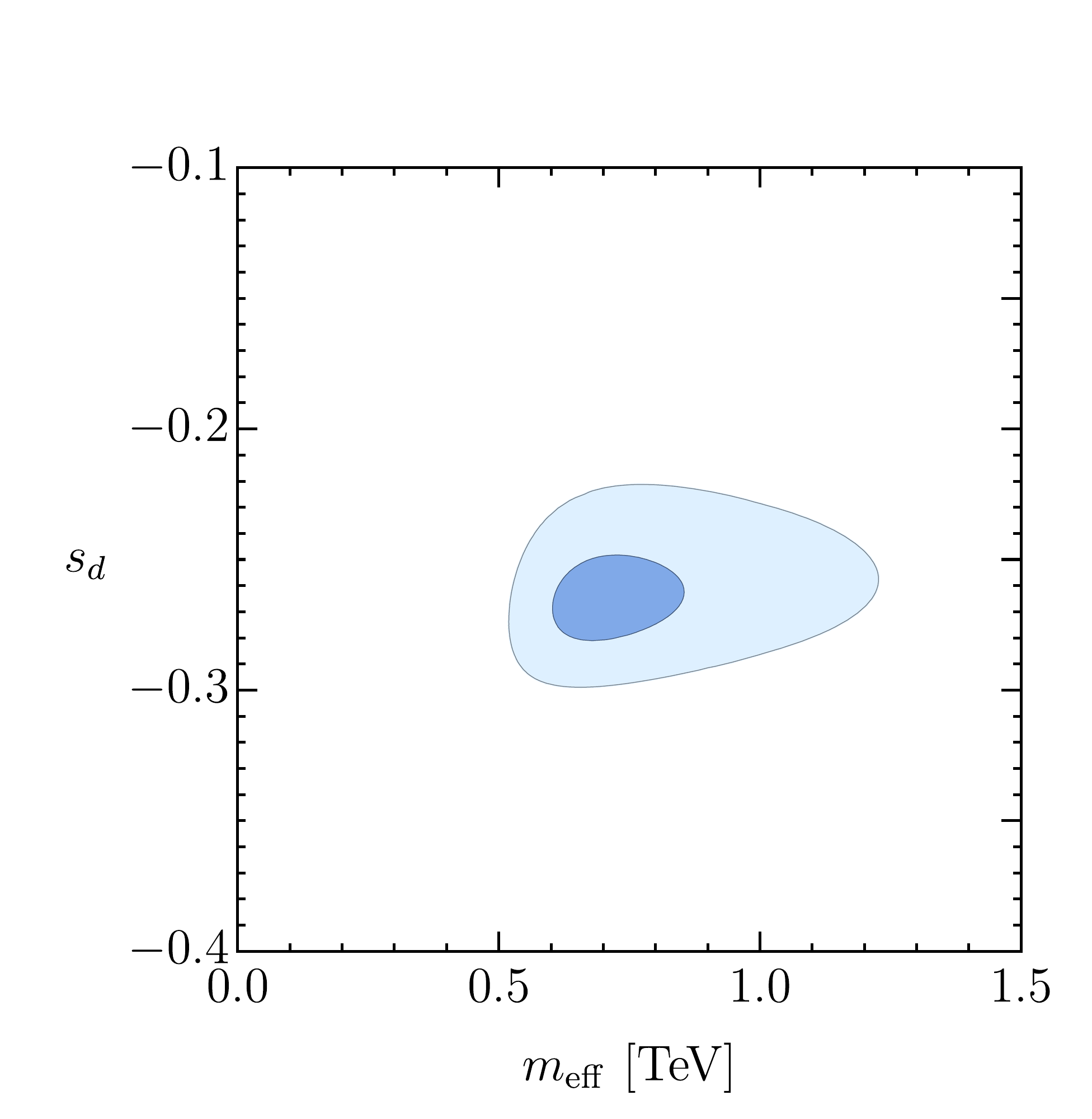}
\includegraphics[scale = 0.35]{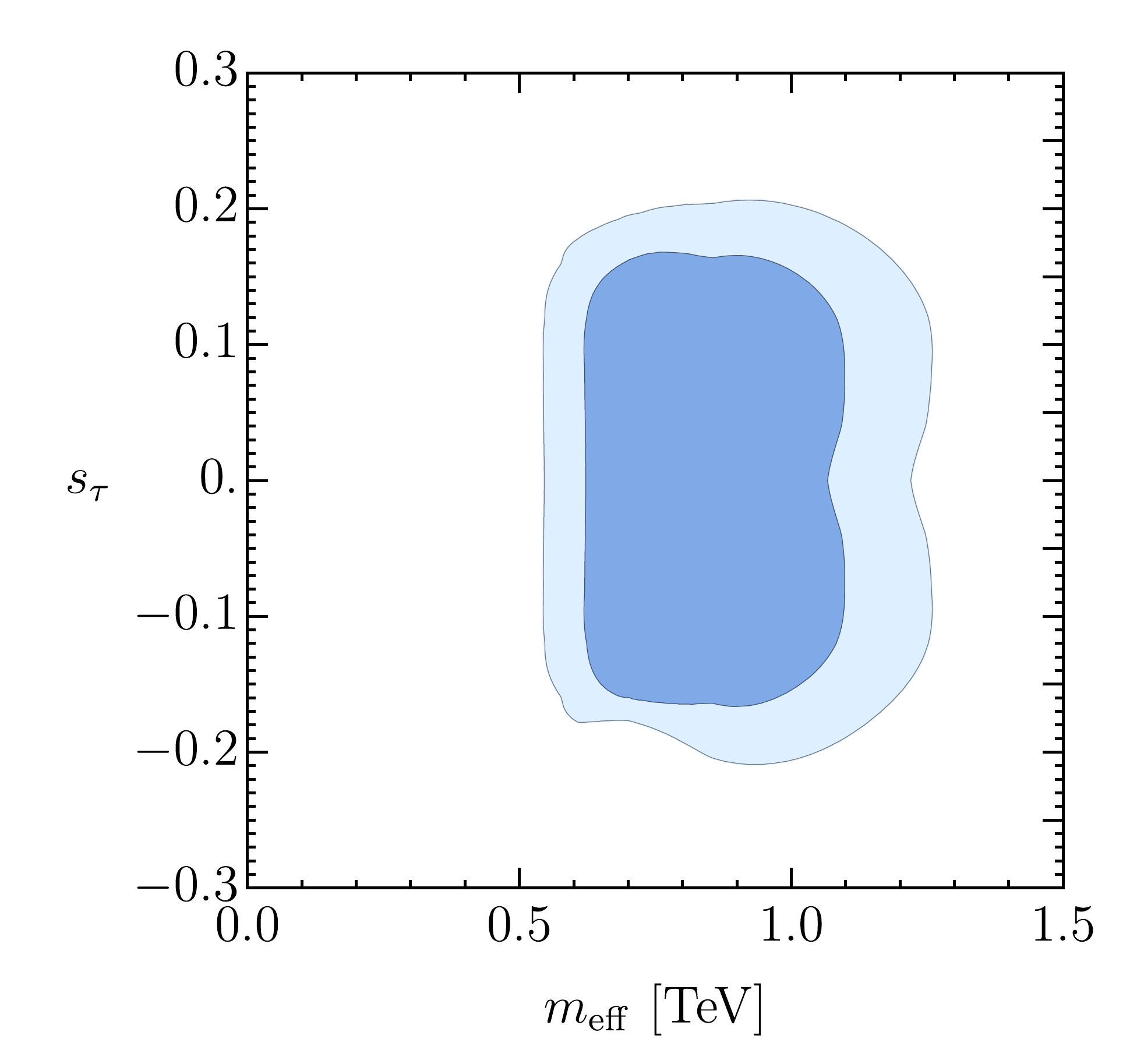}
\includegraphics[scale = 0.35]{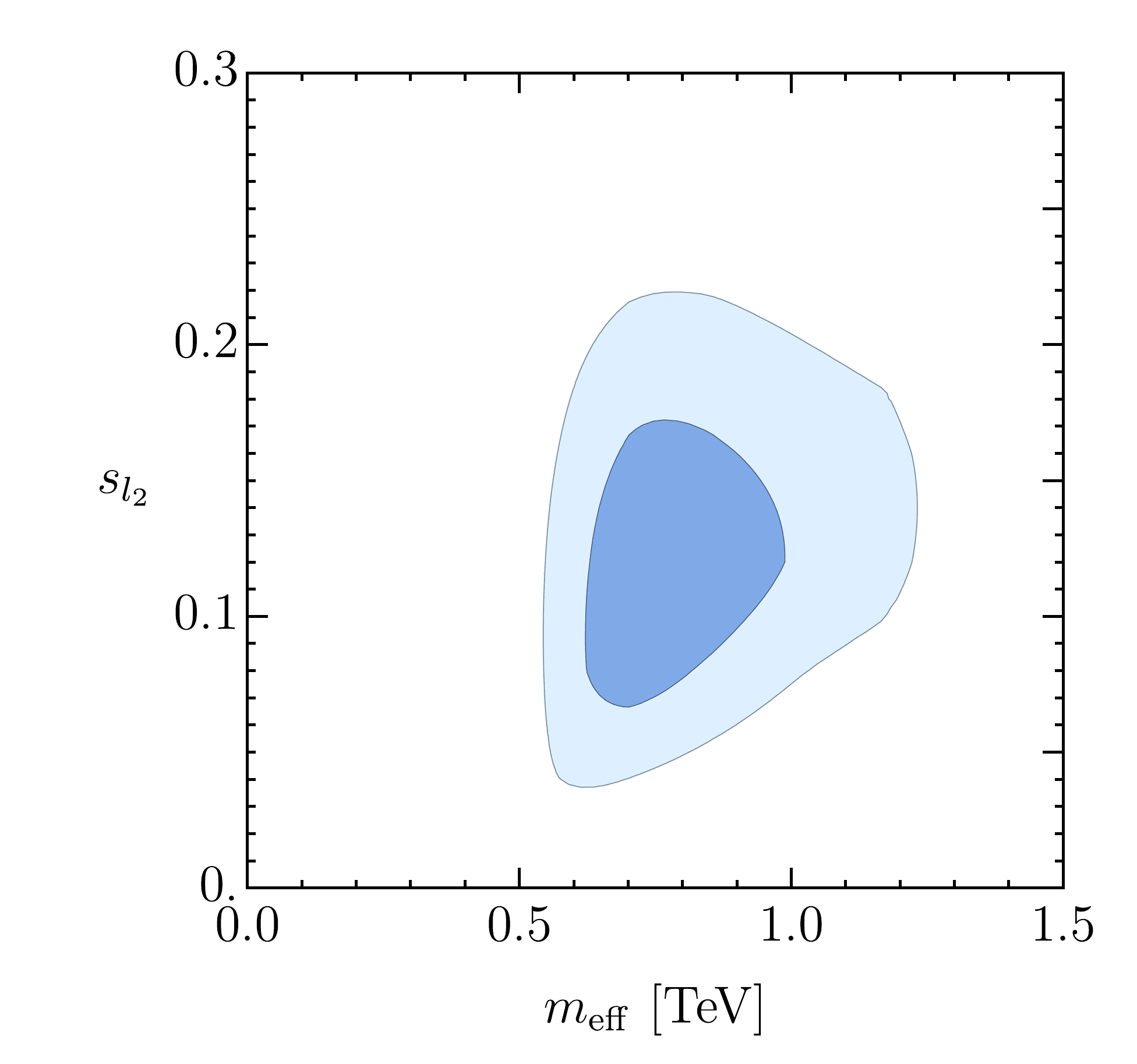}
\caption{Two-dimensional preferred 1 and 2 $\sigma$ regions (in blue and light blue, respectively) for $m_{\rm eff}$ vs $s_b$, $s_d$, $s_\tau$, $s_{l_2}$ for Model 3.
}
\label{fig:2dplots}
\end{figure}

A clear summary of the all these features is provided by the $\delta R_{D^{(\ast)}}$ vs.~$m_{\rm eff}$ plot in Figure \ref{fig:RDmeff_allcases}.
As can be seen, only Model 3 is able to generate a contribution to  $\delta  R_{D^{(\ast)}}$ above $10\%$, provided $m_{\rm eff}$ 
lies close to its lower bound. In this figure, we also show the bound on $m_{\rm eff}$ obtained by searching for modifications of the 
Drell-Yan process $pp\to\tau^+\tau^- +X$ at high-energies, which is sensitive 
to the $t$-channel $U_1$ exchange~\cite{Faroughy:2016osc,Schmaltz:2018nls,Baker:2019sli}. 
A series of recent analyses by CMS of $pp\to\tau^+\tau^- +X$~\cite{CMS:2022a,CMS:2022b}, and the related charged-current process
$pp\to\tau\nu +X$~\cite{CMS:2022c}, allows us to set the bound $m_{\rm eff} \gsim 700$~GeV, 
with a tantalizing $3\sigma$ excess for $m_{\rm eff} \sim 800$~GeV~\cite{CMS:2022b}.

Focusing on Model 3, in Figure~\ref{fig:2dplots} we illustrate the range of the mixing parameters determined by the fit. These plots provide a 
qualitative indication of the degree of flavour alignment necessary to successfully fit both sets of anomalies and, at the same time, 
satisfy all available constraints. As it can be seen, all parameters are compatible with their natural size, hence with the assumption of 
a mildly broken $U(2)^n_f$ symmetry.  However, both  $s_d$ and  $s_b$ require a $10\%$  tuning with respect to their natural sizes:
$s_d=O(\lambda)$ and $s_b = O(|V_{cb}|)$.

In summary, the proposal of a $U_\mu^a$ leptoquark, with couplings to 
fermions ruled by a mildly broken $U(2)^n_f$ flavour symmetry connected to the structure of the SM Yukawa couplings, 
originally formulated in~\cite{Barbieri:2015yvd}, remains a very interesting option to address one or both sets of 
$B$ anomalies. The embedding of the $U_\mu^a$ in the adjoint of $SU(4)$, which is necessary for any realistic UV completion,  
makes this construction more constrained but still viable. In particular, addressing both sets of anomalies is possible only under 
rather specific conditions about the $U(2)^n_f$  symmetry breaking. 
The evolution of the experimental data in the near future will tell if (some of) the anomalies will persist and, 
in the positive case, in which direction their explanation in terms of $SU(4)$ vector leptoquarks will have to evolve.
On general grounds, the relatively low value of $m_{\rm eff}$ that we found in all the simplified models 
we have considered makes the search for high-energy signatures of the leptoquark
(and possibly the vector-like fermions)
quite interesting in view of the high-luminosity phase of the LHC.

\section*{Acknowledgements}
This project has received funding from the European Research Council (ERC) under the European Union's Horizon 2020 research and innovation program under grant agreement 833280 (FLAY), and by the Swiss National Science Foundation (SNF) under contract 200020\_204428. The research of C.C. was supported by the Cluster of Excellence \textit{Precision Physics, Fundamental Interactions, and Structure of Matter} (PRISMA$^+$ - EXC 2118/1) within the German Excellence Strategy (project ID 39083149). C.C. is grateful for the hospitality of Perimeter Institute where part of this work was carried out. Research at Perimeter Institute is supported in part by the Government of Canada through the Department of Innovation, Science and Economic Development Canada and by the province of Ontario through the Ministry of Economic Development, Job Creation and Trade. This research was also supported in part by the Simons Foundation through the Simons Foundation Emmy Noether Fellows Program at Perimeter Institute.

\appendix

\section{Simplified expressions for low-energy observables}
\label{app:obs} 

We denote the relative variation of an observable $O$ with respect to the SM by $\delta(O)$, with 
\begin{align}
\delta(O) = \frac{O - O_{\mathrm{SM}}}{O_{\mathrm{SM}}}\,. 
\end{align}
\subsubsection*{$b \to s ll $ and $b \to c l \nu$}
\begin{equation}
C_{9,\mathrm{NP}}^{\mu} = - C_{10,\mathrm{NP}}^{\mu}  = - \frac{2 \pi}{\alpha V_{ts}^\ast V_{tb}} \left( C_U  \beta^{U}_{s \mu}  \beta^{U \ast}_{b \mu}   -\frac{C_X}{4}  \beta^{X}_{sb} \beta^{X}_{\mu \mu} \right)  
\label{eq:C9} 
\end{equation}  
\begin{equation}
\delta R_{D^{(\ast)}}^{\tau/\mu} =  2  C_U \mathrm{Re} \left[ \frac{\beta^{U \ast}_{b \tau} \beta^{U }_{c\nu_\tau}-\beta^{U \ast}_{b \mu} \beta^{U}_{c\nu_\mu}}{V_{cb}} \right]
\approx 2  C_U \mathrm{Re} \left[  \frac{ V_{cb} \beta^{U \ast}_{b \tau} \beta^{U}_{b \tau} + V_{cs} \beta^{U \ast}_{b \tau} \beta^{U}_{s \tau}   }{ V_{cb}  } \right]
 \label{eq:RDRDs}
\end{equation}

\subsubsection*{Universality tests in leptonic $\tau$ decays}
\begin{align}
\left(\frac{g_\alpha}{g_\beta}\right)_{\ell}=\left[\frac{\mathcal{B}(\ell_\alpha \to \ell_\rho\, \nu\bar\nu)_{\rm exp}/\mathcal{B}(\ell_\alpha \to \ell_\rho\, \nu\bar\nu)_{\rm SM}}{\mathcal{B}(\ell_\beta \to \ell_\rho\, \nu\bar\nu)_{\rm exp}/\mathcal{B}(\ell_\beta \to \ell_\rho\, \nu\bar\nu)_{\rm SM}} \right]^{\frac{1}{2}}\,.
\end{align}
\begin{align}
\left( \frac{g_\tau}{g_\mu} \right)_\ell  & =  1  +  \frac{9}{12} C_X  \left( \abs{ \beta^X_{\tau e}}^2 - \abs{ \beta^X_{\mu e}}^2 \right)  -  \eta  \,  C_U \left(    \abs{ \beta^{U}_{b\tau}}^2    - \abs{ \beta^{U}_{b\mu}  }^2  \right)\\
\left( \frac{g_\tau}{g_e} \right)_\ell  & =  1  +  \frac{9}{12} C_X  \left( \abs{ \beta^X_{\tau \mu}}^2 - \abs{ \beta^X_{\mu e}}^2 \right)  -  \eta  \,  C_U \left(    \abs{ \beta^{U}_{b\tau}}^2    - \abs{ \beta^{U}_{b e}  }^2  \right)  
\end{align}
Neglecting the $X$ contribution and using the fact that   $\abs{ \beta^{U}_{b \mu,e} } \ll \abs{ \beta^{U}_{b \tau} }$, we have:
\begin{align}
\left( \frac{g_\tau}{g_{\mu}} \right)_\ell  \approx \left( \frac{g_\tau}{g_{e}} \right)_\ell \approx 1 - \eta \, C_U \abs{ \beta^{U}_{b\tau}}^2\,,
\label{eq:tauLFUV}
\end{align}
where the running $\eta = 0.079$ is computed assuming $\Lambda = 2 \, \mathrm{TeV}$. 

\subsubsection*{$\tau-\mu$ LFV}

\begin{equation}
\mathcal{B}(\tau \to 3 \mu)  =  \frac98 C_X^2  \abs{\beta_{\mu \tau}^{X} \beta_{\mu\mu}^{X} }^2 \mathcal{B}(\tau \to \mu \bar \nu \nu)_{\mathrm{SM}}
\label{eq:tau23mu}    
\end{equation}

\subsubsection*{$\Delta F =2$ observables}

Effective Lagrangian for the $\Delta F=2$ mixing amplitudes:
\begin{align}
\begin{aligned}
   \mathcal{L}^{\Delta F=2} 
   = - C_{bs} \left( \bar b_L \gamma_\mu s_L \right)^2
    - C_{bd} \left( \bar b_L \gamma_\mu d_L \right)^2
    - C_{uc} \left( \bar u_L \gamma_\mu c_L \right)^2
    + \text{h.c.} \,,
    \label{eq:DF2gen}
\end{aligned} 
\end{align}
NP contribution to these Wilson coefficients:
\begin{align}
\begin{aligned}
 C_{bs}^{\mathrm{NP}} & = \frac{2}{v^2}\left( \frac{C_G}{3}  \left. \beta^{G}_{bs}\right.^2 + \frac{C_X}{24}  \left. \beta^{X}_{bs}\right.^2  \right) \\ 
  C_{ds}^\mathrm{NP}  &=   \frac{2}{v^2}\left( \frac{C_G}{3} \left. \beta^{G}_{ds}\right.^2 + \frac{C_X}{24}  \left. \beta^{X}_{ds}\right.^2 \right)  
   \\
C_{uc}^\mathrm{NP} &=  \frac{2}{v^2}\left( \frac{C_G}{3} \left. \beta^{G,u}_{uc}\right.^2 + \frac{C_X}{24} \left.  \beta^{X,u}_{uc}\right.^2 \right) 
\label{eq:DF2}
  \end{aligned} 
\end{align} 

Main observables are $\mathrm{Im}[C_{uc}^\mathrm{NP}]$, $\mathrm{Im}[C_{ds}^\mathrm{NP}]$ and $\delta (\Delta m_{B_s})$. The latter is defined as

\begin{equation}
  \delta (\Delta m_{B_s})  = \left| 1 + \frac{C^{\rm NP}_{bs}}{C_{bs}^{\rm SM}} \right| - 1  \,,
  \label{eq:DmBs}
   \end{equation}    
   
 where 
   
   \begin{equation}
   C^{\rm SM}_{bs} = \frac{G_F^2 m_W^2}{4\pi^2}\,\big( V_{tb}^* V_{ts} \big)^2\,S_0(x_t)  \,, 
\end{equation}

with  $x_t = \frac{m_t^2}{m_W^2}$ and $ S_0(x_t)\approx 2.37$. 
\subsubsection*{$\mu-e$ LFV}
\begin{align}
\begin{aligned}
\mathcal{B}(K_L^0 \to \mu^- e^+) & = \frac{\tau_{K_L} G_F^2 f_{K}^2 m_\mu^2   m_K}{8 \pi}   \left( 1 - \frac{m_\mu^2}{m_K^2}\right)^2     \abs{ C_U \left( \beta^{U}_{de}  \beta^{U \ast}_{s\mu}      +    \beta^{U}_{se}  \beta^{U \ast}_{d\mu}  \right) - \frac{ C_X}{4}   2  \mathrm{Re}\left[ \beta^{X}_{ds}\right]  \beta^X_{\mu e}   }^2  \\
\mathcal{B}(K_L^0 \to \mu^+e^-) & = \frac{\tau_{K_L} G_F^2 f_{K}^2 m_\mu^2 m_K }{8 \pi  }      \left( 1 - \frac{m_\mu^2}{m_K^2}\right)^2     \abs{   C_U \left(  \beta^{U}_{d\mu} \beta^{U \ast}_{se}  + \beta^{U}_{s\mu}  \beta^{U \ast}_{de} \right)    -  \frac{C_X}{4}  2  \mathrm{Re}\left[ \beta^{X}_{ds}\right]  \beta^X_{e\mu}     }^2
 \end{aligned}
\end{align}

\begin{align}
\mathcal{B}(\mu \to 3 e)    &=   \frac98  C_X^2 \abs{\beta_{e \mu}^X \beta_{ee}^X }^2 \mathcal{B}(\mu \to e \bar \nu \nu)_{\mathrm{SM}} 
\end{align}

\bibliographystyle{JHEP}
\bibliography{paper}

\end{document}